\documentstyle[emulateapj]{article}
\begin{document}

\title{Radial Temperature Profiles of 11 Clusters of Galaxies
Observed With {\it BeppoSAX}}

\author{Jimmy A. Irwin\altaffilmark{1} and Joel N. Bregman}
\affil{Department of Astronomy, University of Michigan, \\
Ann Arbor, MI 48109-1090 \\
E-mail: jirwin@astro.lsa.umich.edu, jbregman@umich.edu}

\altaffiltext{1}{Chandra Fellow.}

\begin{abstract}

We have derived azimuthally-averaged radial temperature profiles
of the X-ray gas contained within 11 clusters of galaxies with redshift
$z=0.03-0.2$ observed with
{\it BeppoSAX}. Each of the 11 clusters have had their radial temperature
profiles previously determined with {\it ASCA}. We find that the temperature
profiles of these clusters are generally flat or increase slightly out to
$\sim$30\%
of the virial radius, and that a decline in temperature of 14\% out to
30\% of the virial radius is ruled out at the 99\% confidence level. This is
in accordance with a previous {\it ROSAT} PSPC study
and an {\it ASCA} study by White (1999),
but in disagreement with an {\it ASCA} study by Markevitch et al.\ (1998)
that found on average
that cluster temperature profiles decreased significantly with radius.

\end{abstract}


\section{Introduction} \label{sec:intro}

Knowledge of the radial temperature profile of the hot gas contained within
galaxy clusters is a crucial element in determining the total gravitational
mass of clusters. Through the equation of hydrostatic equilibrium, the
total mass of the cluster can be derived if the density gradient, temperature,
and temperature gradient of the gas is known. The latter quantity is the
most difficult quantity to obtain, and it has generally been assumed that
the gas is isothermal in most previous approaches, since collimated X-ray
instruments such as {\it Ginga} and {\it EXOSAT} could not determine the
temperature structure of clusters.

The assumption of isothermality of the hot gas has been called into question in
recent years, mainly as a result of studies done with {\it ASCA}. The ability
of {\it ASCA} to perform spatially-resolved spectroscopy over the 1--10 keV
energy range made it the first X-ray instrument capable of addressing the
issue of temperature structure in hot clusters. Many {\it ASCA} studies have
found that the gas within clusters is not isothermal, but decreases with
increasing radius, in some cases up to a factor of two (e.g., Markevitch 1996;
Markevitch et al.\ 1998; Markevitch et al.\ 1999). However, other studies of
clusters using {\it ASCA} data have come to the conclusion that the
gas is largely isothermal (e.g., White 1999; Fujita et al.\ 1996;
Ohashi et al.\ 1997;
Kikuchi et al.\ 1999), at least outside of the cooling radius of cooling
flow clusters. A likely cause of this discrepancy is the handling
of the large, energy-dependent point spread function (PSF) of {\it ASCA}
that preferentially scatters hard X-rays. This creates an artificial
increase in the temperature profile with radius if not dealt with properly.
The PSF-correction method applied by Markevitch et al. consistently leads to
significantly decreasing temperature profiles, while other methods (most
notably the method of White 1999) lead to isothermal profiles.

The discrepancy among the different PSF-correction methods prompted an
analysis of {\it ROSAT} PSPC data by Irwin, Bregman, \& Evrard (1999). Although
{\it ROSAT} was only sensitive to photon energies up to 2.4 keV and was
therefore not the most ideal instrument with which to study hot clusters,
large (factor of two) differences in temperature should have been detected,
but were not. The composite X-ray ``color" profiles for 26 clusters in
the Irwin et al.\ (1999) survey indicated isothermality outside of the
cooling radius. In fact, a 20\% temperature drop within 35\% of the virial
radius was ruled out at the 99\% confidence level.

In this paper, we attempt to resolve the temperature profile discrepancy
using {\it BeppoSAX} data. {\it BeppoSAX} is sensitive to photon energies
up to 10.5 keV, and has a half-power radius that is one-half that of the
{\it ASCA} GIS instrument. In addition, the PSF of {\it BeppoSAX} is only
weakly dependent on energy. Thus, {\it BeppoSAX} is better-suited 
for determining temperature profiles for clusters of galaxies than previous
X-ray telescopes. Using a sample
of 11 clusters found in the {\it BeppoSAX} archive, we derive radial
temperature profiles for each cluster. In a future paper, we will discuss
the abundance profiles of the 11 clusters. Throughout this paper,
we assume $H_0=50$ km s$^{-1}$ Mpc$^{-1}$ and $q_0=0.5$.

\section{Sample and Data Reduction} \label{sec:sample}

From the {\it BeppoSAX} Science Data Center (SDC) archive (available at
http://www.sdc.asi.it/sax\_main.html) we have obtained data for seven cooling
flow clusters (A85, A496, A1795, A2029, A2142, A2199, and 2A0335+096)
and four non-cooling
flow clusters (A2163, A2256, A2319, and A3266). All the clusters have redshifts
in the range $z=0.03-0.09$ except A2163 ($z=0.203$). We analyze data from
the Medium Energy Concentrator Spectrometer (MECS), which consists of
three identical gas scintillation proportional counters (two detectors after
1997 May 9) sensitive in the 1.3--10.5 keV energy range. A detailed
description of the MECS is given in Boella et al.\ (1997). The event files
were subjected to the standard screening criteria of the {\it BeppoSAX}
SDC.

Since our goal is spatially-resolved spectroscopy, accounting for scattering
from the PSF of the MECS is important. As stated above, scattering from the
PSF has an enormous impact on the temperature profiles derived from
{\it ASCA} data. Fortunately, the detector + telescope
PSF of {\it BeppoSAX} is nearly independent of energy, unlike
the {\it ASCA} GIS. This is because the Gaussian PSF of the MECS detector
improves with increasing energy, while the PSF of the grazing incidence Mirror
Unit degrades with increasing energy (D'Acri, De Grandi, \& Molendi 1998),
leading to a partial cancellation when these two effects are combined.
Still, it is important to account for the PSF accurately when deriving
temperature profiles.

To correct for the PSF we have used the routine {\it effarea}, available
as part of the SAXDAS 2.0 suite of {\it BeppoSAX} data reduction programs.
This routine is described in detail in Molendi (1998) and
D'Acri et al.\ (1998). Briefly, {\it effarea} creates an appropriate
effective area file that corrects for vignetting and scattering effects
for an azimuthally-symmetric circular or annular region. It does so by creating
correction vectors that are a function of energy, and which when multiplied
by the observed spectrum yields the corrected spectrum. The surface brightness
profile of the cluster (determined from the analysis of {\it ROSAT} data by
Mohr, Mathiesen, \& Evrard 1999 and Ettori \& Fabian 1999) is convolved
with the PSF of the MECS in order to determine the extent to which
scattering from other regions of the cluster have contaminated the emission
from the extraction region in question. This information is incorporated
into the auxiliary response file (the {\it .arf} file), which is subsequently
used in the spectral fitting. The correction to the observed spectrum is
modest; D'Acri et al.\ (1998) and Kaastra, Bleeker, \& Mewe (1998) found
only small changes between the uncorrected and corrected temperature profiles
for Virgo and A2199, respectively. The mismatch in energy bandpasses
between the {\it ROSAT} (0.2--2.4 keV) surface brightness profile and
{\it BeppoSAX} (1.65--10.5 keV) does not appear to have significantly
affected the results.

For each cluster, spectra were extracted from concentric annular regions
centered on the peak of emission of the cluster
with inner and outer radii of $0^{\prime}-2^{\prime}$, $2^{\prime}-4^{\prime}$,
$4^{\prime}-6^{\prime}$, and $6^{\prime}-9^{\prime}$. We also extracted one
global spectrum from $0^{\prime}-9^{\prime}$. At $9^{\prime}$ the
telescope entrance window support structure (the strongback) becomes a factor.
In addition, for off-axis angles greater than $10^{\prime}$, the departure
of the PSF from radial symmetry becomes noticeable (Boella et al.\ 1997).
This coupled with the fact that some of the clusters have poor photon
statistics outside of $10^{\prime}$ prompted us to end our profiles at
$9^{\prime}$. At this radius, our temperature profiles extend out to 55\% of
the virial radius, $r_{virial}$, for A2163 and 17\%--33\% for the other
clusters, where $r_{virial} = 3.9~(T/10~{\rm keV})^{1/2}$ Mpc. Background
was obtained from the deep blank sky data provided by the SDC. We used the
same region filter to extract the background as we did the data, so that
both background and data were affected by the detector response in the same
manner. The energy channels were rebinned to contain at least 25 counts.

The procedure outlined above does not fit the spectrum of the various
regions within the cluster simultaneously. Instead, it assumes a uniform
spectrum throughout when correcting for contamination from other regions.
Whereas this is not important for the innermost bin (since very few photons
are scattered in from larger radii compared to the number of photons truly
belonging in the innermost bin), this might affect the outer bins if
the temperature profile is varying strongly. This does not seem to be the
case though. The spectrum correction vectors presented in D'Acri et al.\ (1998)
for A2199 were quite modest. Other than their innermost bin (which loses
some flux via scattering but does not receive much scattered flux from
exterior bins), the corrections amounted to 5\% or less for energies above
3 keV. In addition, if the exterior bins were significantly contaminated
by emission from the interior of the cluster that was at a considerably
different temperature, it is likely that no single-component thermal model would
give an adequate fit to the data. However, all the spectral fits of the third
and fourth spatial bins of the clusters in our study were adequate, with
nearly all fits having $\chi_{\nu} < 1.05$. This coupled with the fact that
the PSF-corrected temperatures were not significantly different from the
uncorrected temperatures (see \S~\ref{ssec:temp_profiles}) indicates
that our results are not strongly affected by not fitting the spectra
from different regions simultaneously.

For one of the lower temperature clusters (A2199) a long pointed {\it ROSAT}
PSPC observation was available in the HEASARC archive for which no
{\it ROSAT}-determined temperature profile had been published. The
observation (RP800644N00) was filtered such that all time intervals with
a Master Veto Rate above 170 counts s$^{-1}$ were excluded, in order to
discard periods of high background. This resulted in a net exposure of
34,232 seconds. Background was taken from an annulus with inner and outer
radii of $30^{\prime}$ and $40^{\prime}$. The background was scaled to and
subtracted from the source spectra, which were subsequently binned such that
each energy channel contained at least 25 counts. Energy channels below
0.2 keV were ignored in the fit.

\begin{figure*}[htb]
\vskip4.1truein
\includegraphics{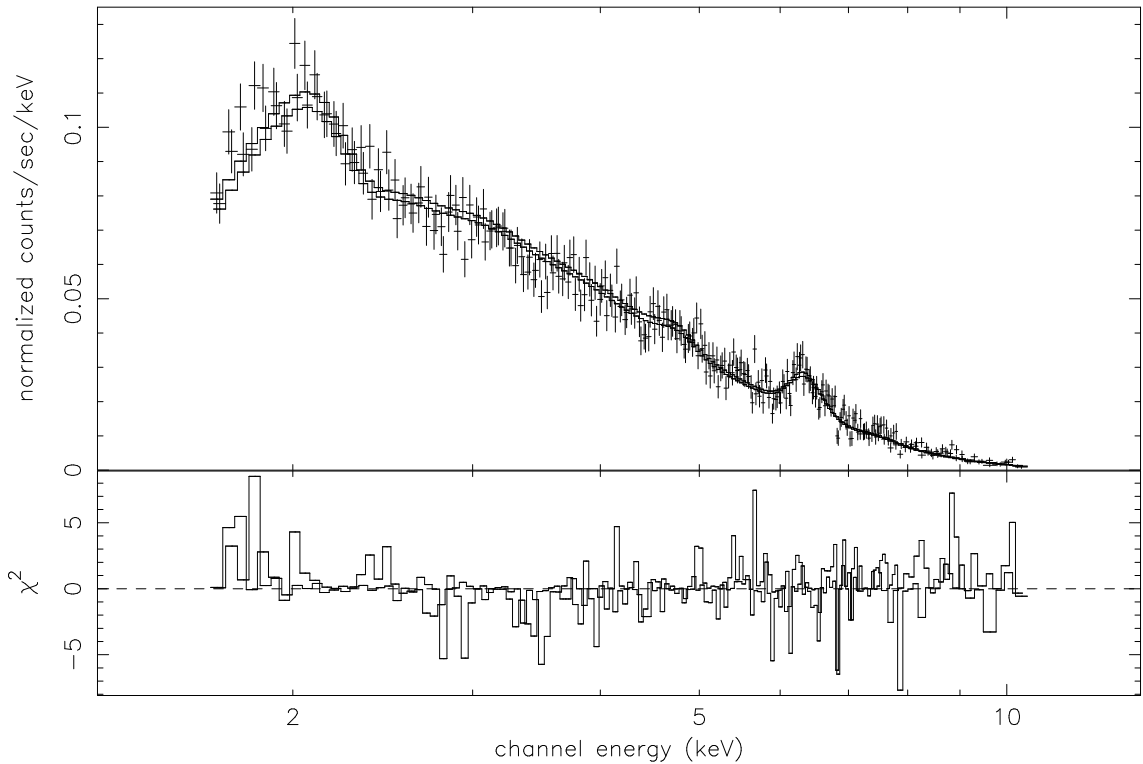}
\caption[A2256 Spectrum]{
Best-fit MEKAL model spectrum with residuals to the global spectrum of the
non-cooling flow
cluster A2256 using data in the 1.65--10.5 keV range. An excess of positive
residuals below 3.0 keV was a feature common in all 11 clusters analyzed
here.
\label{fig:a2256}}
\end{figure*}

\section{Temperature of the Hot Gas} \label{sec:temps}

\subsection{Global Temperatures} \label{ssec:temp_global}

Using XSPEC we used a MEKAL model with an absorption component fixed at the
Galactic value in all spectral fits, except the {\it ROSAT} PSPC observation
of A2199. The temperature, metallicity, and
redshift were allowed to vary. Spectral fits of {\it BeppoSAX} MECS data
of the Perseus cluster with the redshift fixed showed significant residuals in
the iron line complex region around 6.7 keV (R. Dupke 1999, private
communication). These residuals disappeared when the redshift was allowed to
vary. The cause of this feature is a systematic shift of 45--50 eV in the MECS
channel--to--energy conversion (F. Fiore 1999, private communication).
This systematic shift was evident in our sample; when the redshift was
allowed to vary, the measured redshift was less than the optically-determined
redshift in all 11 clusters, and inconsistent with the optically-determined
redshift at the 90\% confidence level for eight of them. A modest decrease in
the reduced $\chi^2$ also occurred for most of the clusters when the redshift
was allowed to vary. However, freeing the redshift did not affect the values
obtained for the temperature and metallicity significantly (less than a
5\% change in either quantity).

With the model described above, we fit the data from MECS2 and MECS3 (and
MECS1 when available) separately, but with the same normalization. In
accordance with the {\it Cookbook for BeppoSAX NFI Spectral Analysis} energy
channels below 1.65 keV and above 10.5 keV were ignored in the fit. On the
whole, the fits to the global spectra were rather poor, ranging from
$\chi^2_{\nu}=1.13-1.8$ for 170--556 degrees of freedom.
Inspection of the residuals revealed that the poor
fits resulted from excess emission in the 1.65--3.0 keV range. The best-fit
spectrum from the non-cooling flow cluster A2256 is shown in
Figure~\ref{fig:a2256} and illustrates the positive residuals below 3.0 keV.
This effect was even more pronounced in the cooling flow clusters. We
performed the fits again, this time only using data in the 3.0--10.5 keV
range. The fits were much better, with the fits to all clusters having
$\chi^2_{\nu} \le 1.20$. In addition, the global temperature values for the
fits performed in the 3.0--10.5 keV range were much closer to the global
values determined from {\it ASCA} data (see Table~\ref{tab:compare}).
Ten of the 11 temperatures derived from the 1.65--10.5 keV fit were below
the {\it ASCA} value, and in eight cases the 90\% error bars did not overlap.
Conversely, nine of the 11 temperatures derived from the
3.0--10.5 keV fit have 90\% errors bars that overlap with the
error bars from the {\it ASCA} temperatures.
All quoted
errors are 90\% confidence levels unless otherwise noted.

Since the improvement in $\chi^2_{\nu}$ when channels in the 1.65--3.0 energy
range were excluded was more pronounced for the cooling
flow clusters than in the non-cooling flow clusters, we investigated the
possibility that the excess emission below 3.0 keV was a result of a
cooling flow component. Indeed, the addition of a cooling flow model
to the MEKAL model provided a substantially better fit in the
1.65--10.5 keV case, and in some cases the fit became formally acceptable.
However, the inferred cooling rates were several hundred solar masses per year
higher than previous published values (e.g., Peres et al. 1998). In fact,
cooling rates of several hundred solar masses per year were found for the
non-cooling flow clusters A2163, A2256, A2319, and A3266. In addition,
significant cooling rates were found for spectra extracted from regions of the
cluster far from the cluster center, where no cooling gas should be found. This
clear contradiction illustrates how a physically implausible model can still
yield good spectral fits, and the danger in interpreting such a result. We
conclude that although some of the excess emission in the 1.65--3.0 keV range
is from cooling gas, there is a clear excess of soft emission beyond
what is expected from cooling gas, possibly due to uncertainties in the
calibration of the MECS instruments. Since including this energy range leads
to global temperatures significantly below the {\it ASCA} value, we
only fit the data in the 3.0--10.5 keV range for the remainder of the paper.
Given the good agreement in the 3.0--10.5 keV fit and the {ASCA}-determined
temperatures, we are confident of the calibration of {\it BeppoSAX} above
3.0 keV.

\begin{table*}[htb]
\caption[Global Fits]{}
\begin{center}
\begin{tabular}{ccccccccc}
\multicolumn{9}{c}{\sc Global Temperature Fits} \cr
\tableline \tableline
&& {\it ASCA}\tablenotemark{a} &&
\multicolumn{2}{c}{\it BeppoSAX} 1.65--10.5 keV &
& \multicolumn{2}{c}{\it BeppoSAX} 3.0--10.5 keV \cr
\cline{5-6} \cline{8-9}
Cluster && $kT$ (keV) && $kT$\tablenotemark{b} (keV) & $\chi_{\nu}$/d.o.f. &
& $kT$\tablenotemark{b} (keV) & $\chi_{\nu}$/d.o.f. \\
\tableline
A85  && $6.1\pm0.2$&& 5.6$^{+0.1}_{-0.1}$& 1.70/340 && 6.4$^{+0.3}_{-0.2}$ & 1.17/282 \\
A496 && $4.3\pm0.2$&& 3.7$^{+0.1}_{-0.1}$& 1.59/317 && 4.2$^{+0.1}_{-0.1}$ & 1.20/259 \\
A1795&& $6.0\pm0.3$&& 5.0$^{+0.2}_{-0.2}$& 1.14/269 && 6.0$^{+0.4}_{-0.4}$ & 0.97/211 \\
A2029&& $8.7\pm0.3$&& 6.7$^{+0.2}_{-0.2}$& 1.26/170 && 7.6$^{+0.5}_{-0.4}$ & 0.98/141 \\
A2142&& $8.8\pm0.6$&& 7.6$^{+0.2}_{-0.2}$& 1.32/350 && 8.7$^{+0.4}_{-0.4}$ & 1.02/292 \\
A2163&&11.5&&11.0$^{+0.6}_{-0.6}$& 1.32/556 && 11.7$^{+1.0}_{-0.9}$ &1.05/440 \\A2199&& $4.4\pm0.2$&& 4.0$^{+0.1}_{-0.1}$& 1.58/506 && 4.4$^{+0.1}_{-0.1}$ & 1.03/419 \\
A2256&& $7.5\pm0.4$&& 6.2$^{+0.3}_{-0.3}$& 1.13/296 && 7.1$^{+0.5}_{-0.4}$ & 1.06/238 \\
A2319&& $9.2\pm0.7$&& 8.8$^{+0.3}_{-0.4}$& 1.27/331 && 10.5$^{+0.8}_{-0.7}$ &1.04/273 \\
A3266&& $7.7\pm0.8$&& 8.0$^{+0.4}_{-0.3}$& 1.15/323 && 9.9$^{+0.8}_{-0.7}$ & 0.93/265 \\
2A0335+096&&$\sim$3.4\tablenotemark{c}&&2.80$^{+0.03}_{-0.04}$&1.80/315&&3.20$^{+0.08}_{-0.08}$
& 1.13/257 \\
\tableline
\end{tabular}
\end{center}
\tablenotetext{a}{Single-fit temperature from Markevitch et al.\ (1998).}
\tablenotetext{b}{Errors listed are 90\% confidence levels.}
\tablenotetext{c}{Estimated value inside of 10$^{\prime}$ from radial
temperature profile of Kikuchi et al.\ (1999).}
\label{tab:compare}
\end{table*}

\subsection{Radial Temperature Profiles} \label{ssec:temp_profiles}

\begin{figure*}[htb]
\vskip4.6truein
\includegraphics{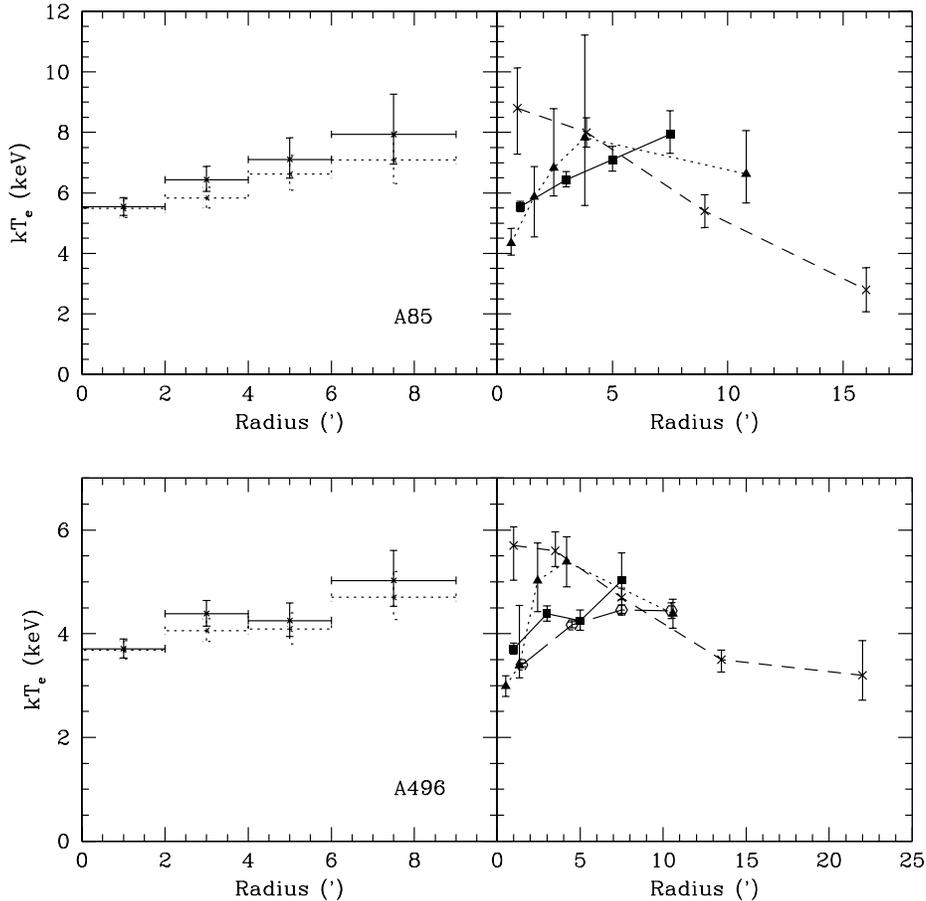}
\caption[Temp Profiles]{
Radial temperature profiles for the 11 clusters in the sample, derived from
fitting in the 3.0--10.5 keV range. In the left panel, solid lines represent
temperatures corrected for the PSF of {\it BeppoSAX} and the dotted lines
are not corrected. Error bars are 90\% confidence levels. The right panels
compare our results (filled squares and solid lines) with temperature profiles
derived from {\it ASCA} data using the method of White (1999; filled triangles
and dotted line), the method of Markevitch et al.\ (1996; crosses and dashed
lines), uncorrected {\it ASCA} profiles from Dupke \& White (1999) and
Kikuchi et al.\ (1999) (open hexagons and long dashed
line in A496 and 2A0335+096), and {\it BeppoSAX} profiles for A2029, A2199,
A2319, and
A3266 (open triangles and long dashed lines). For the right panels, the error
bars are 1$\sigma$ confidence levels.
\label{fig:temp_profiles}}
\end{figure*}

The PSF-corrected and -uncorrected radial temperature profiles for each of the
11 clusters are shown in the left panels of
Figure~\ref{fig:temp_profiles}. As was found in previous
studies, correction for the MECS PSF does not seriously affect the
temperature profile. In the right panels are our {\it BeppoSAX} temperature
profiles along with temperature profiles derived from other {\it ASCA}
and {\it BeppoSAX} studies for comparison. We note that for A85, A496,
A1795, A2029, A2142, and A2199 the innermost bin has been fit with a cooling
flow component in addition to a thermal model for the profiles of
Markevitch et al.\ (1998), Markevitch et al.\ (1999),
and Sarazin, Wise, \& Markevitch  (1998), which accounts for
the large discrepancy in this bin compared to other studies that fit the
spectra with only single-component models.

{\it A85}: We have included the subclump to the south of the cluster center
in our analysis. Outside of the cooling radius, the temperature
profile increases moderately from 6.4 keV to 7.9 keV at a
significance level of $2.2\sigma$. This is in agreement
with the {\it ASCA} analysis by White (1999), but contrasts with
Markevitch et al.\ (1998) who found a profile that decreased from 8.0 keV
in a $1\farcm5-6^{\prime}$ annular bin to 6.3 keV in a
$6^{\prime}-12^{\prime}$ annular bin with {\it ASCA} data.
Pislar et al.\ (1997) and Kneer et al.\ (1995) analyzed {\it ROSAT} PSPC
data for this cluster and found a roughly isothermal profile outside the
cooling region. However, the {\it ROSAT} analysis found a significantly
lower temperature, with most of the cluster below 5 keV. This is possibly
due to a gain calibration problem of the {\it ROSAT} PSPC instrument.

{\it A496}: This nearby cluster ($z=0.0326$) has a cooling rate of
$\dot M = 95~M_{\odot}$ yr$^{-1}$ (Peres et al.\ 1998).
The temperature is lowest in the
center (typical of a cooling flow), and is consistent with a constant
value of 4.5--5.0 keV at larger radii. This is consistent with the {\it ASCA}
analysis of Dupke \& White (1999), who did not perform a PSF correction.
For low temperature clusters, the PSF does not introduce a significant
spurious positive gradient to the temperature profile (Takahashi et al.\ 1995).
The profile of White (1999) also agreed for the most part with our profile,
although one of their bins deviated by $\sim2\sigma$ from ours.
Markevitch et al.\ (1999) found a continuous drop in temperature
of 5.6 keV to 3.5 keV from a
$2^{\prime}-5^{\prime}$ annular bin to a $10^{\prime}-17^{\prime}$ annular bin.

{\it A1795}: We find a temperature profile consistent with a temperature
of 6--7 keV outside of the cooling flow region. A significant decrease in
temperature was not found in the center for this cooling flow cluster. This
result is similar to the {\it ASCA} results of Mushotzky et al.\ (1995;
uncorrected for the PSF), White (1999), and Ohashi et al.\ (1997),
although the Ohashi et al.\ (1997) result found a somewhat lower overall
temperature for this cluster.
Markevitch et al.\ (1998) detected a decline in temperature with
radius, but not at a high significance with {\it ASCA}.
{\it ROSAT} found a low temperature in the inner cooling flow region,
and a moderately increasing profile at larger radii (Briel \& Henry 1996).
The discrepancy in the innermost bin is likely the result of the low energy
bandpass (0.1--2.4 keV) of {\it ROSAT} sampling lower temperature gas than
the higher energy bandpass (3.0--10.5 keV) of {\it BeppoSAX} in the cooling
region.

{\it A2029}: We find a basically flat temperature profile out to $6^{\prime}$ 
and a marginally significant ($1.7\sigma$) rise from $6^{\prime}-9^{\prime}$,
consistent with the {\it ASCA} analysis of White (1999). This is
the opposite trend found by Sarazin et al.\ (1998) with
{\it ASCA} data, who found a decline in
temperature from $\sim$9 keV to 6 keV outside of $5^{\prime}$ (but with large
errors in the outer regions). Still, they found that an isothermal profile was
rejected at the $>96\%$ confidence level. This was one of the few clusters for
which Irwin et al.\ (1999) found evidence for a statistically significant
temperature decline with {\it ROSAT}
data, although the drop did not occur until outside of 10$^{\prime}$.
Molendi \& De Grandi (1999) analyzed the same
{\it BeppoSAX} data and found a temperature profile consistent with 8 keV
out to 8$^{\prime}$ and dropping to 5 keV from $8^{\prime}-12^{\prime}$ (see
\S~\ref{ssec:bepposax} for a detailed comparison of the two analyses).

{\it A2142}: We find a very flat profile with a temperature of 8--9 keV.
A similar result was found by White (1999), although the errors were large.
Markevitch et al.\ (1998)
found evidence for a temperature decline, but not at a high significance level.
Henry \& Briel (1996) analyzed {\it ROSAT} data and found temperatures
of 10 keV or higher outside the cooling region, with a peak in the
$2\farcm5-5^{\prime}$ bin.

{\it A2163}: This cluster has the highest redshift in our sample ($z=0.203$),
and is also the hottest. The cluster does not possess a cooling flow and
probably underwent a merger in the recent past
(e.g., Elbaz, Arnaud, \& B\"ohringer 1995). Since the centroid of the cluster
lies over $5^{\prime}$ from the detector center, we have excluded data
from behind the strongback support structure and beyond. We have included
data in the last annular
bin out to $12^{\prime}$ (as long as it fell inside the strongback) to
improve the statistics in the last bin. Our profile extends out to 73\%
of the virial radius, considerably farther than any other cluster in our
sample. We find that the cluster has a temperature of 10--11 keV out to
4$^{\prime}$, before experiencing a marginally significant ($<2\sigma$)
rise in temperature at larger radii. The errors are quite large at large
radii, although the temperature is greater than 8 keV at the 90\% level.
This agrees with the {\it ASCA} result of White (1999),
but disagrees strongly with the result from {\it ASCA} and {\it ROSAT} by
Markevitch et al.\ (1996), who found temperatures of 12.2$^{+1.9}_{-1.2}$,
11.5$^{+2.7}_{-2.9}$, and 3.8$^{+1.1}_{-0.9}$ for annular bins of
$0^{\prime}-3^{\prime}$, $3^{\prime}-6^{\prime}$, and $6^{\prime}-10^{\prime}$
in extent, respectively. For these regions, we find temperatures of
10.1$^{+0.9}_{-0.8}$, 11.7$^{+2.6}_{-1.8}$, and 13.2$^{+17.9}_{-4.6}$,
respectively, using only data inside the strongback (90\% errors). Thus,
the outermost bin of the {\it BeppoSAX} data differs from that of
Markevitch et al.\ (1998) at the 3.3$\sigma$ confidence level.

{\it A2199}: Outside of the cooling flow region, we find a constant
temperature of about 4.5 keV. Analysis of the same {\it BeppoSAX} data by
Kaastra et al.\ (1998) and of {\it ASCA} data by White (1999) found a similar
result. However, Markevitch et al. (1999) found a
steadily decreasing profile from 5.2 keV to under 5 keV at $7\farcm5$, with
a further decline at larger radii with the same {\it ASCA} data.

We have analyzed the {\it ROSAT} PSPC data for this cluster, and derived
a temperature profile out to $18^{\prime}$ (0.9 Mpc). The angular extent
of this cluster is large, so to confirm that our background region was
not significantly contaminated by the cluster emission, we derived a
temperature

\begin{figure*}[htb]
\vskip4.4truein
\includegraphics{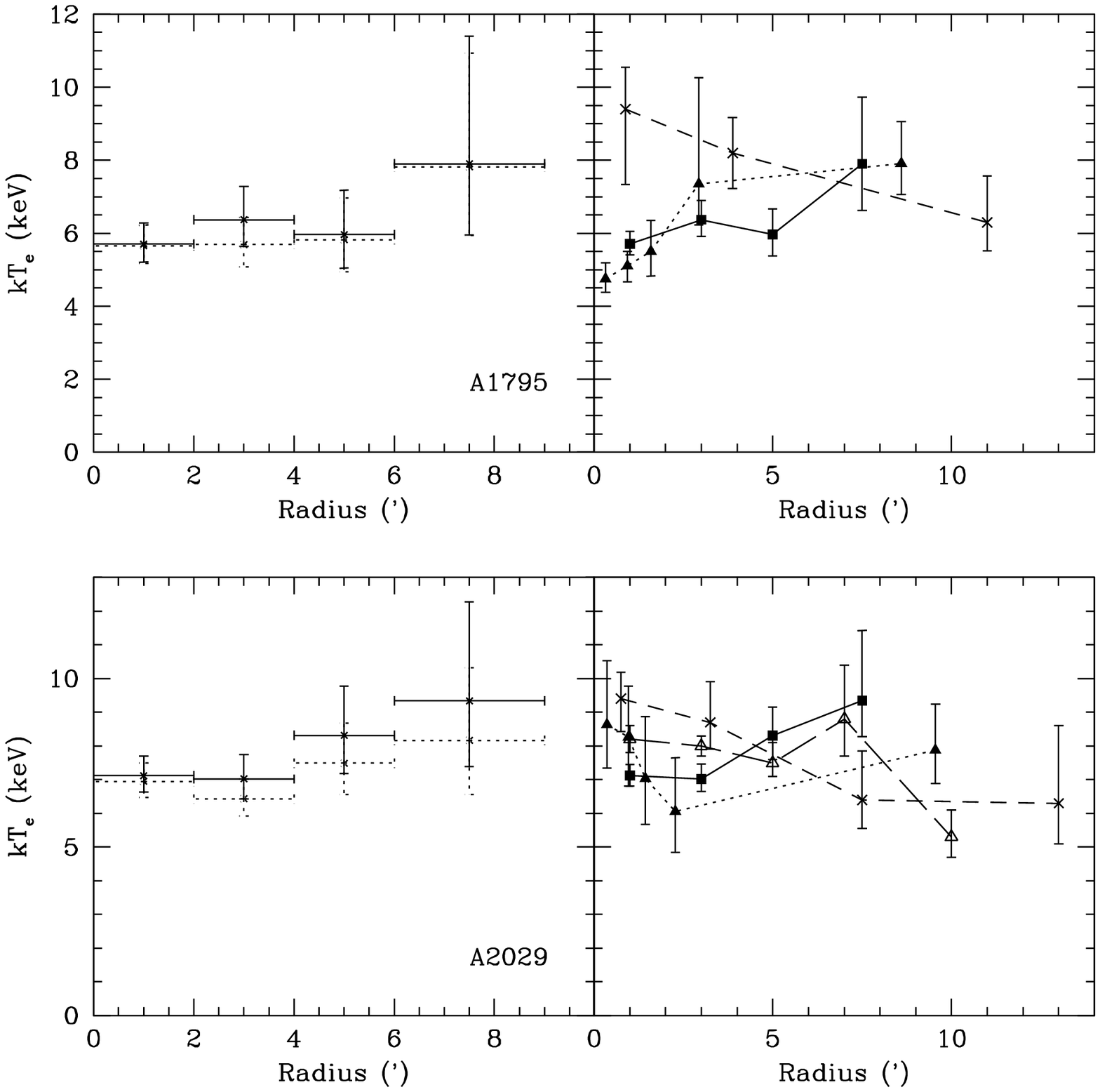}
\end{figure*}

\begin{figure*}[htb]
\vskip4.64truein
\includegraphics{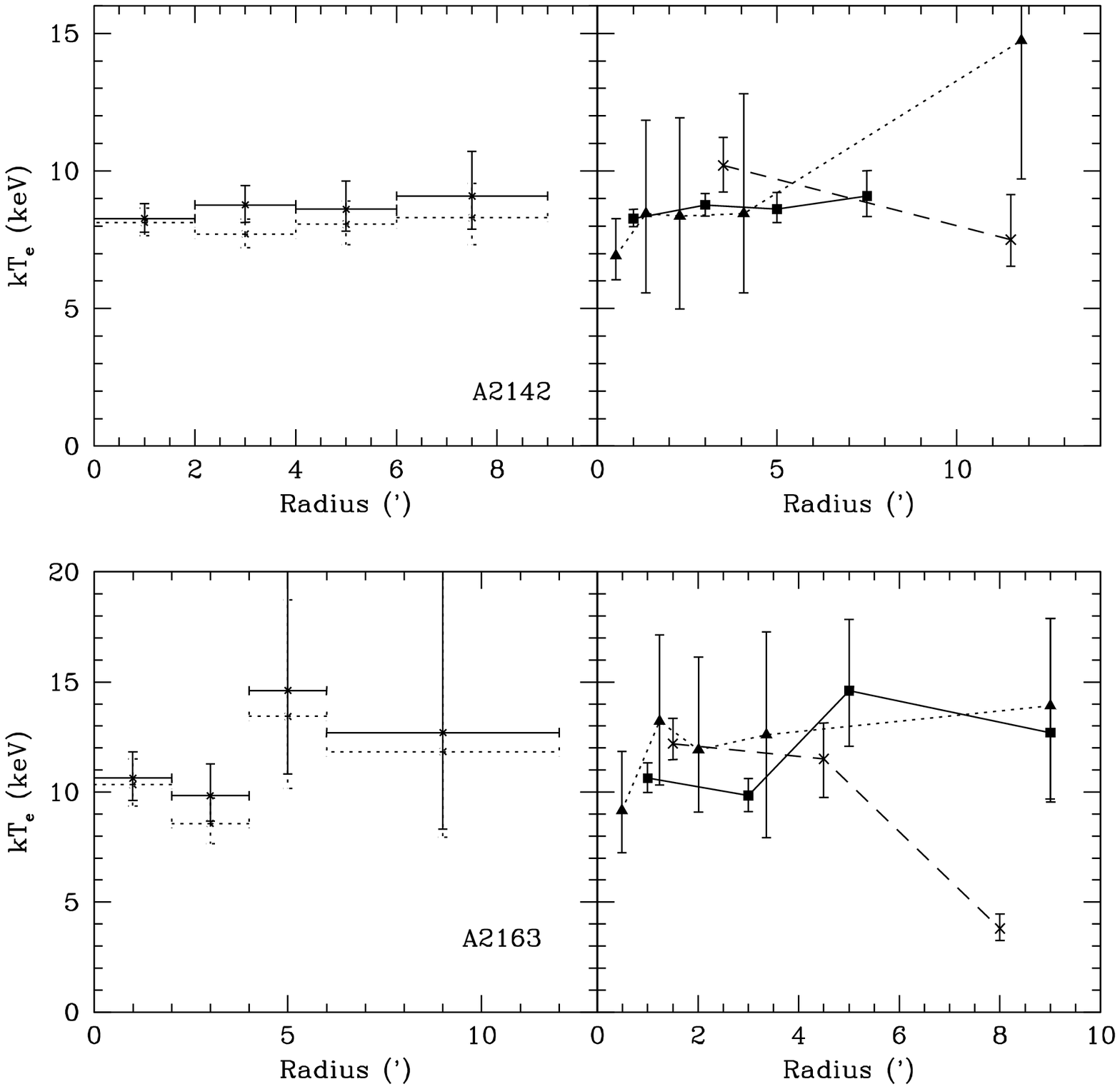}
{
~~~~~~~~~~~~~~~~~~~~~~~~~~~~~~~~~~~~~~~~~~~~~~~~~~~~~~~~~~~~~~~~~~Fig. 2 -- continued.
}
\end{figure*}

\begin{figure*}[htb]
\vskip4.4truein
\includegraphics{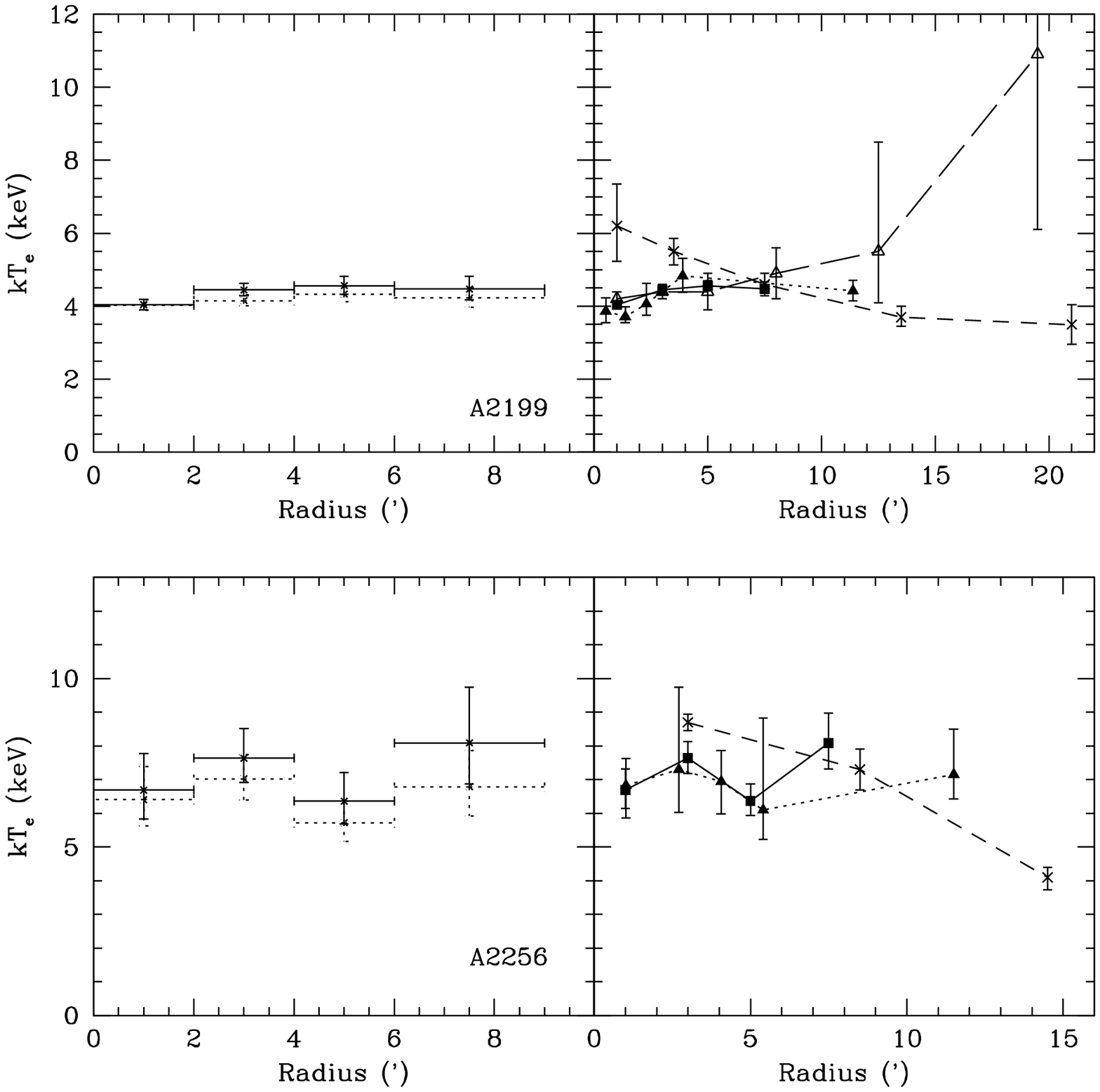}
\end{figure*}

\begin{figure*}[htb]
\vskip4.64truein
\includegraphics{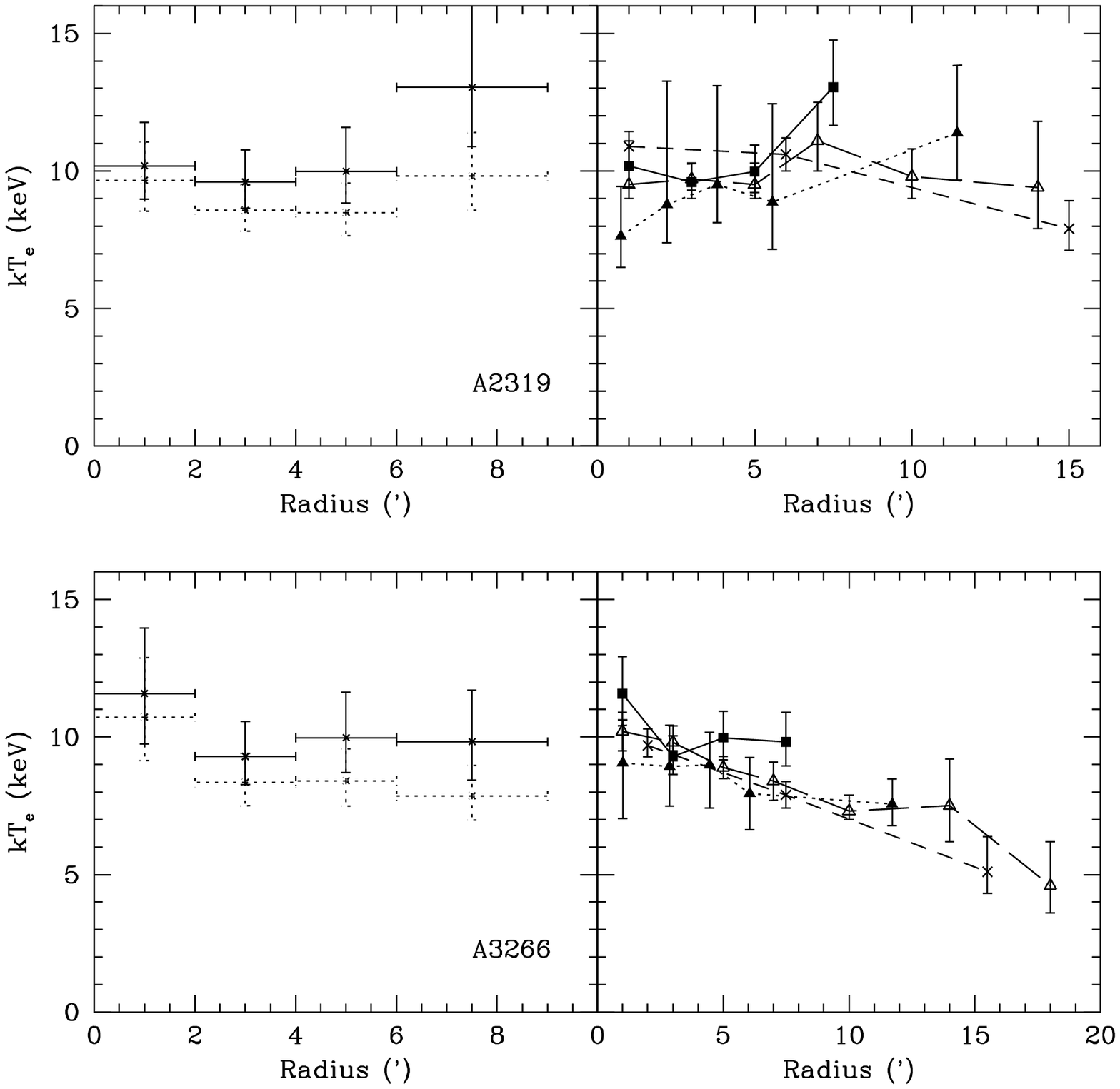}
{
~~~~~~~~~~~~~~~~~~~~~~~~~~~~~~~~~~~~~~~~~~~~~~~~~~~~~~~~~~~~~~~~~~Fig. 2 -- continued.
}
\end{figure*}

\begin{figure*}[htb]
\vskip2.6truein
\includegraphics{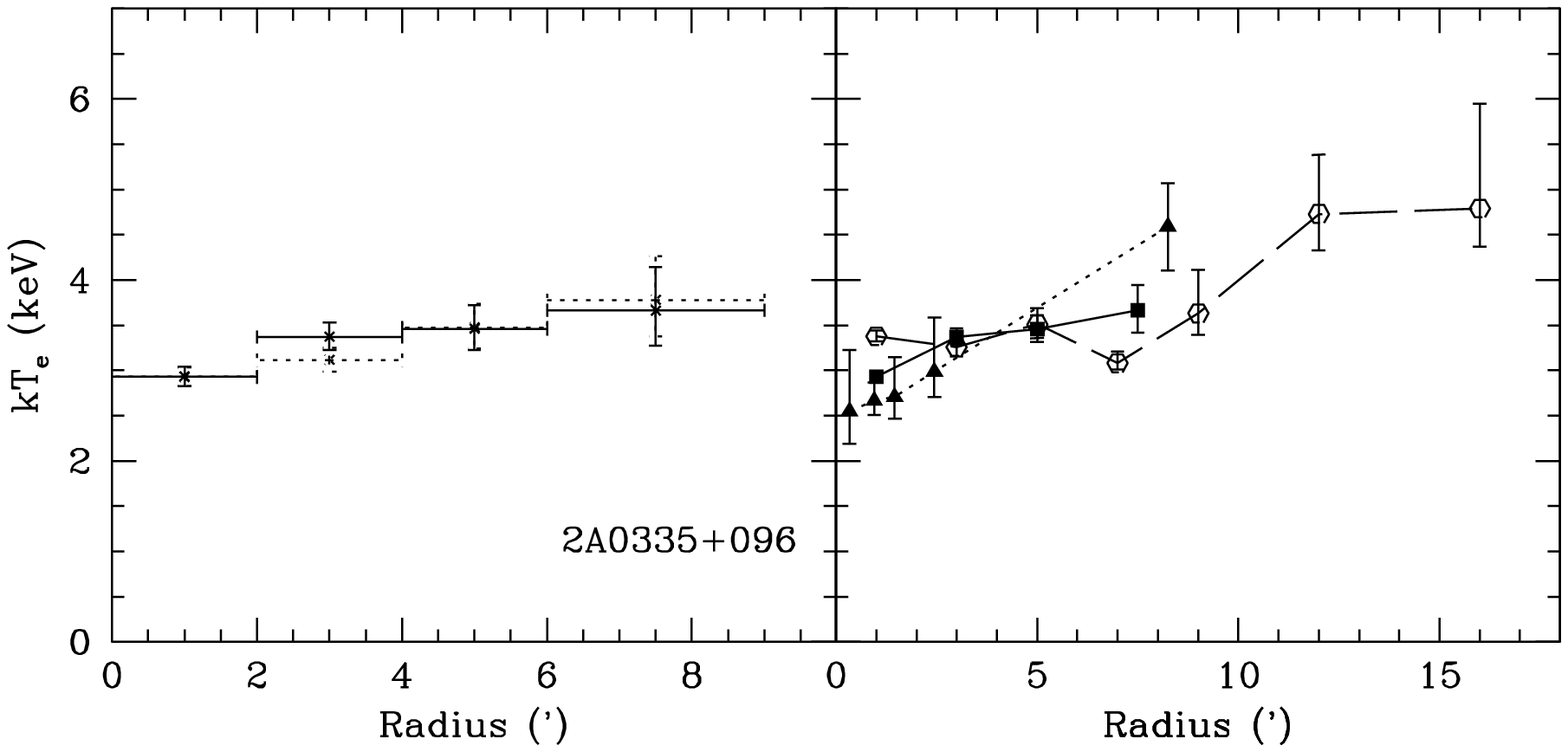}
\vspace{-0.1truein}
{
~~~~~~~~~~~~~~~~~~~~~~~~~~~~~~~~~~~~~~~~~~~~~~~~~~~~~~~~~~~~~~~~~~Fig. 2 -- continued.
}
\end{figure*}
\begin{figure*}[htb]
\vskip4.6truein
\includegraphics{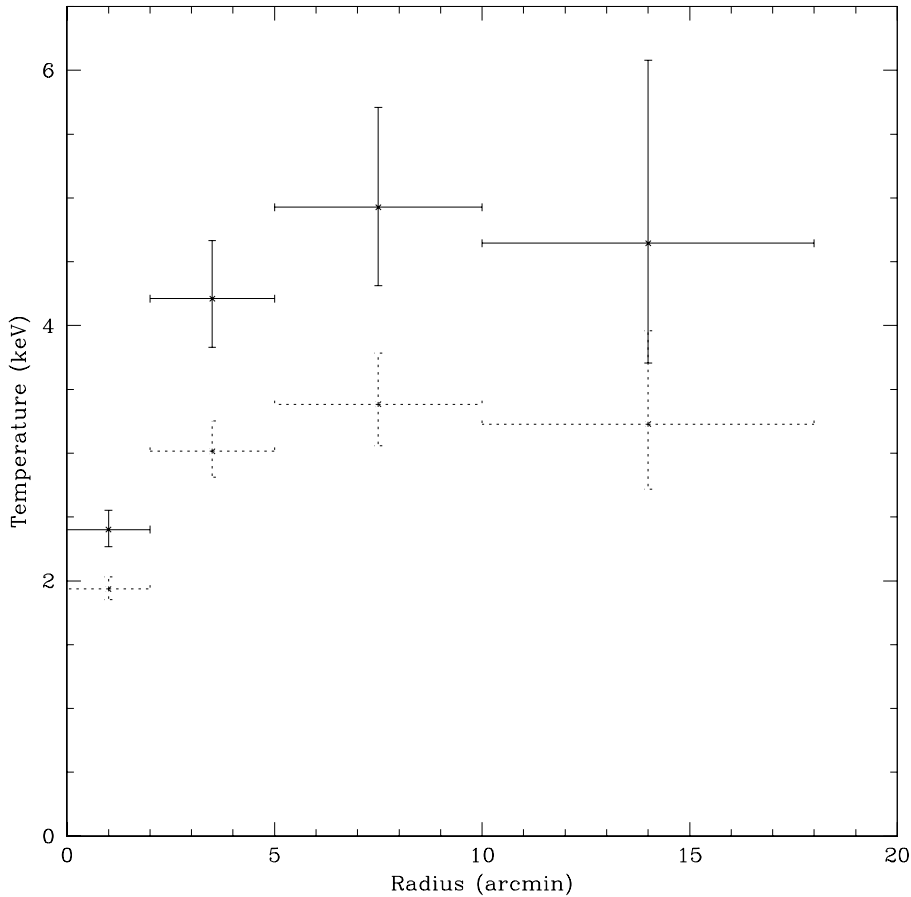}
\caption[A2199]{
Temperature profile of A2199 derived from the {\it ROSAT} PSPC with 90\%
error bars. Outside of the cooling radius, the temperature profile
is flat both in the case where the gain is set to the nominal value
(dotted line) or adjusted such that the temperature in the
$2^{\prime}-9^{\prime}$ region is normalized to the temperature derived from
{\it BeppoSAX} for the same region (solid line).
\label{fig:a2199}}
\end{figure*}

\noindent profile using a background annulus of $40^{\prime}-50^{\prime}$
and found the same results as we did with a background annulus of
$30^{\prime}-40^{\prime}$. The temperature profile is shown in
Figure~\ref{fig:a2199}. The profile shows a drop in the center indicative
of a cooling flow. At large radii the profile is flat. However, the
{\it ROSAT}-derived temperature is significantly lower than the value
from {\it ASCA} or {\it BeppoSAX}. Whereas this might be expected in the
central cooling flow region where {\it ROSAT} is sampling cooler gas than
the other instruments because of its low energy bandpass, this tendency
persists outside the cooling region. From $2^{\prime}-9^{\prime}$, the
temperature is $3.2\pm0.2$ keV, whereas it is $4.5\pm0.1$ for {\it BeppoSAX}.
Data from {\it ROSAT} appears sometimes to have a tendency to measure lower
temperatures than other instruments for clusters, such as A3558
(Markevitch \& Viklinin 1997) and A85 (see above).

To compensate for this effect, we have adjusted the gain of the observation
such that the temperature in the $2^{\prime}-9^{\prime}$ region matched that
of {\it BeppoSAX}. We excluded the inner $2^{\prime}$ to avoid complications
from the cooling flow region. An adjustment of 1.5\% in the gain was necessary
to bring the global temperature determined by the two instruments into
agreement. A gain adjustment of this magnitude was
within the range of values found by Henry \& Briel (1996) when they analyzed
five different pointings of A2142 with {\it ROSAT}. With this new gain value
the temperature profile remains flat outside of the cooling region out to
$18^{\prime}$ (35\% of the virial radius), albeit at a higher value than
before.
The main conclusion drawn from the {\it ROSAT} result of A2199 is that the
temperature profile appears flat outside of the cooling flow region
regardless of whether or not the gain was adjusted.

{\it A2256}: We find a flat temperature profile consistent with a
temperature of 7 keV, in excellent agreement with the {\it ASCA}-determined
profile of White (1999). This contrasts with the Markevitch (1996) result from
{\it ASCA} which found the temperature to decrease from 8.7 keV to 7.3 keV from
the $0^{\prime}-6^{\prime}$ to $6^{\prime}-11^{\prime}$, with a steep
drop to 4 keV at larger radii. Markevitch (1996) claims that the {\it ROSAT}
data confirm this result, albeit with larger errors, contrary to the claim
of Briel, \& Henry (1993) who analyzed the same {\it ROSAT} data
and found a roughly isothermal profile.

{\it A2319}: We find a flat temperature profile out to $6^{\prime}$ and
a marginally significant ($\sim2\sigma$) increase from $6^{\prime}-9^{\prime}$.
This is consistent with the White (1999) result.
Markevitch (1996) also found an isothermal profile out to $10^{\prime}$
with {\it ASCA}, and a decreasing profile at larger radii. The {\it ROSAT}
data suggested isothermality (Irwin et al.\ 1999). Molendi et al.\ 1999
analyzed the same {\it BeppoSAX} data and found a flat temperature profile
out to 16$^{\prime}$.

{\it A3266}: This probable merging cluster exhibits a slight
increase in temperature in the
center and levels off at radii out to 9$^{\prime}$. With {\it ASCA} data,
Markevitch et al.\ (1998) found a steady decrease in temperature from
almost 10 keV in the central $2\farcm5$ to 5 keV outside of
$10^{\prime}$, and White (1999) also found a modestly decreasing profile
(from 9 keV to 7.5 keV). The {\it ROSAT} data suggested isothermality
although a modest decrease in temperature could not be ruled out
(Irwin et al.\ 1999). De Grandi \& Molendi (1999) analyzed the same
{\it BeppoSAX} data and found a more substantial decrease in temperature,
with a drop from 10 keV in the center to 4.5 keV out to 20$^{\prime}$.

{\it 2A0335+096}: The coolest cluster in our sample, 2A0335+096 possesses a
rather strong cooling flow ($400~M_{\odot}$ yr$^{-1}$; Irwin \& Sarazin 1995). 
The temperature is lowest in the innermost bin and levels off to a value of
3.4 keV out to 9$^{\prime}$. Kikuchi et al.\ (1999) analyzed the
{\it ASCA} data for this cluster and found a flat temperature profile out
to 10$^{\prime}$, while White (1999) found a flat profile out to 4$^{\prime}$
and a jump to 4.6 keV at larger radii.

In conclusion, the temperature profiles of the 11 clusters in our sample
are roughly constant, and in general agreement with those derived from
{\it ASCA} data by White (1999). This is in
in disagreement with the profiles derived from the same
{\it ASCA} data with the method of Markevitch et al. (1996), which find
a significant decrease in temperature in many of the clusters. A more detailed
comparison is given in \S~\ref{sec:compare}.

\subsection{Normalized Temperature Profiles} \label{ssec:temp_norm}

\begin{figure*}[htb]
\vskip4.0truein
\includegraphics{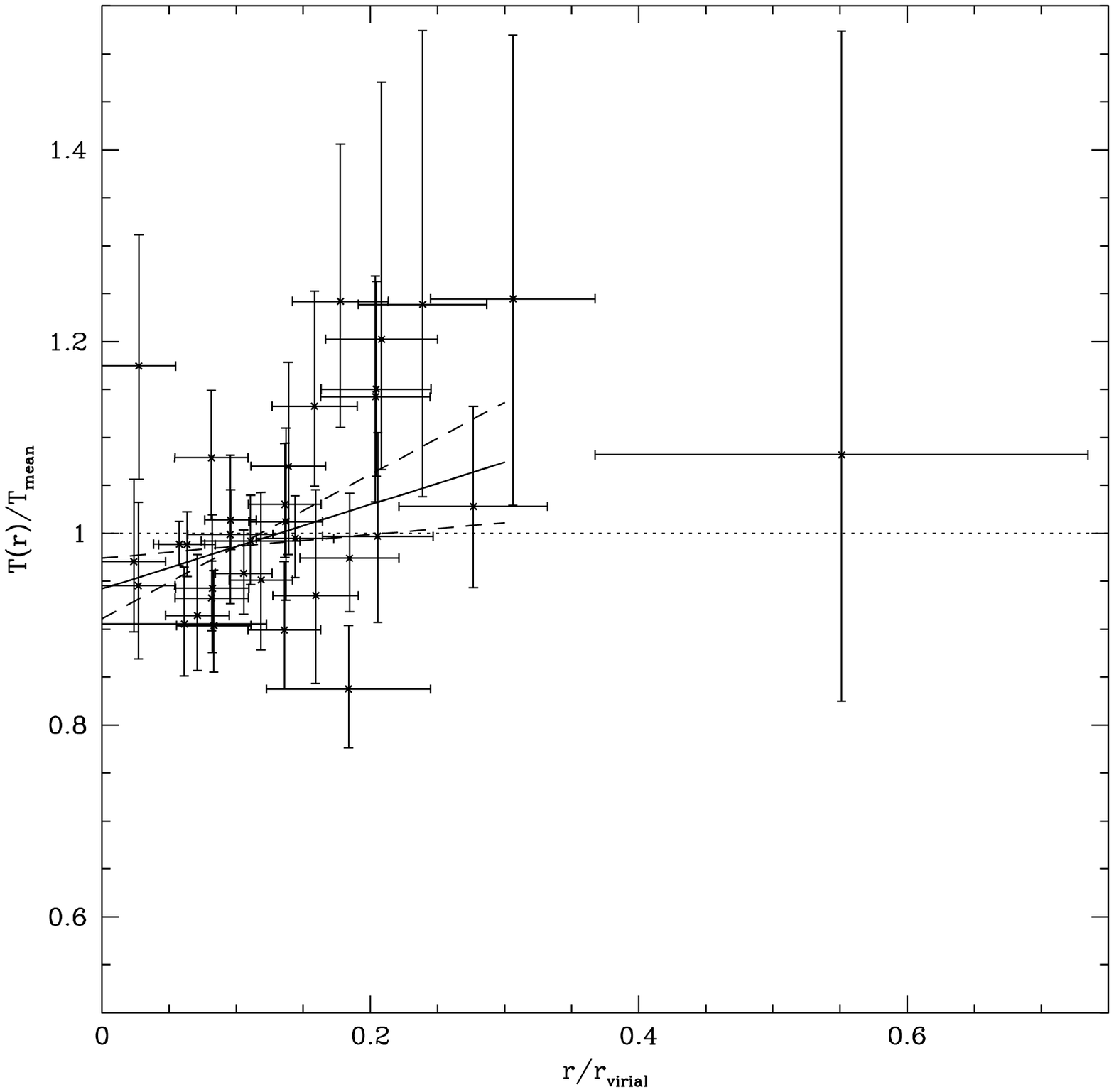}
\caption[t-tmean]{
($a$) Normalized temperature profiles for all 11 clusters in the sample
versus radius in units of the virial radius.
($b$) Normalized temperature profiles after compensating for the effects of
cooling flows (see text). Error bars represent the 1 $\sigma$ errors. The
solid line is the best-fit linear
function, with the dotted lines representing the 90\% confidence levels of
the slope.
\label{fig:t-tmean}}
\end{figure*}

In Figure~\ref{fig:t-tmean}{\it a} we plot the temperature profiles normalized
to the global temperature for all 11 clusters versus radius in units of the
virial radius, $r_{virial}$. The error bars represent the 1 $\sigma$
uncertainties. At small radii, the normalized profiles are
typically less than one, owing to the presence of cooling flows in seven of the
11 clusters. As a result of the low temperature in the center, the outer
regions are necessarily normalized to a value greater than one. To compensate
for this effect, we have calculated global temperatures for the seven cooling
flow clusters excluding the inner $2^{\prime}$, and then normalized the
$2^{\prime}-4^{\prime}$, $4^{\prime}-6^{\prime}$, and $6^{\prime}-9^{\prime}$
bins by this temperature, while excluding the innermost bin. The result is
shown in Figure~\ref{fig:t-tmean}{\it b}. Out to 20\% of the virial radius,
the temperature profiles appear flat. From 20\% to 30\% of the virial radius,
the profiles rise somewhat, although the temperatures are not well
constrained in this region. Most of the values are consistent with
unity at the 1$\sigma$  confidence level.

A constant temperature model ($T/T_{mean}=1$) provided a good 
fit to the data ($\chi^2 = 43.7$ for 37 degrees of freedom). A linear model
of the form $(T/T_{mean})=a+b(r/r_{virial})$ provided a somewhat better fit
($\chi^2 = 36.0$ for 35 degrees of freedom), with values of $a=0.942$ and
$b=0.440$. The 90\% confidence range on the slope $b$ was 0.123--0.752.
This best-fit line is shown in Figure~\ref{fig:t-tmean}{\it b}, with the
dashed lines representing the 90\% confidence levels on the slope of the
line. We find that a temperature drop of 14\% from the center out to 30\% of
the virial radius can be ruled out at the 99\% confidence level.

The data indicate that the gas is isothermal or mildly increasing in
temperature out to 30\% of the virial radius (1.0$h_{50}^{-1}$ Mpc for a 7 keV
cluster). Beyond this radius, the temperature profile may well decline. Most
hydrodynamical cluster simulations predict a drop in temperature past 50\% of
the virial radius (see, e.g., Frenk et al.\ 1999). {\it XMM} will be the ideal
instrument for determining the temperature profiles of clusters at large radii.

\begin{figure*}[htb]
\vskip4.0truein
\includegraphics{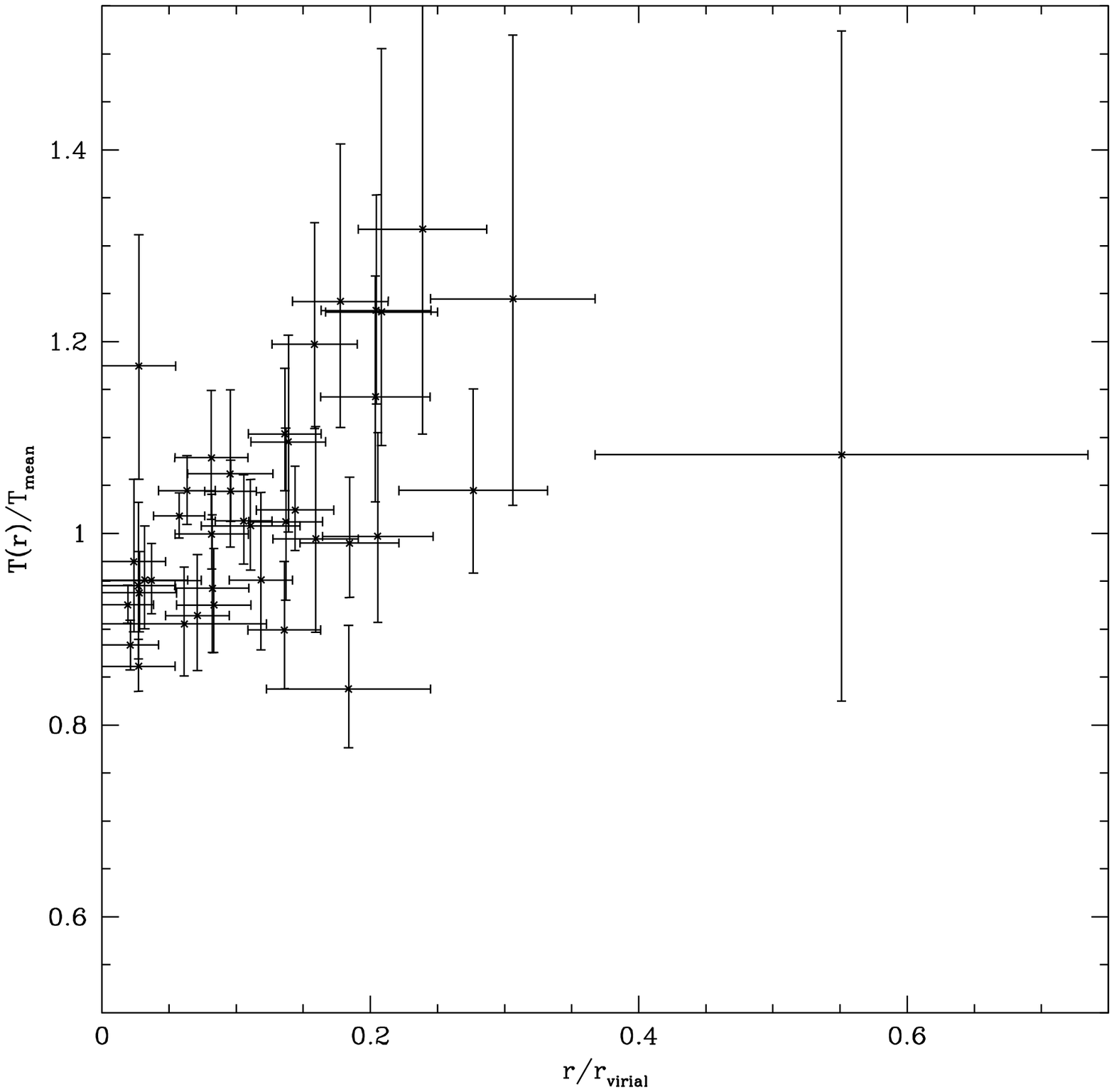}
\end{figure*}

\section{Comparison to Other Results} \label{sec:compare}

The presence of temperature gradients in clusters of galaxies has been a
topic of hot debate in recent years, and there have been many investigations
as to the temperature structure of clusters. Here, we summarize previous
results grouped by X-ray telescope and compare them to our results.

\subsection{Temperature Profiles Determined With {\it ASCA}} \label{ssec:asca}

The claim of the existence of large temperature declines with radius was
first suggested following the results from {\it ASCA} data using the
PSF-correction technique of Markevitch et al.\ (1996), including
Markevitch (1996), Markevitch et al.\ (1998), Sarazin et al.\ (1998), and
Markevitch et al.\ (1999). Markevitch et al.\ (1998) observed 30 clusters
with {\it ASCA} and found in general that the temperature profiles
declined with radius (up to a factor of two), and that the decline was
described well by a polytropic equation of state, i.e.,
\begin{equation}
T(r) \propto \left( 1 + \frac{r^2}{a_x^2} \right)^{-3\beta (\gamma-1)/2} ,
\end{equation}
where $r$ is the projected distance, $a_x$ is the core radius, $\beta$
has its usual meaning in the context of isothermal $\beta$-models, and
$\gamma$ is the polytropic index. The best-fit polytropic index was
$\gamma=1.24^{+0.20}_{-0.12}$ (90\% confidence levels). This model
clearly does not fit
the {\it BeppoSAX} data. The polytropic fit led to $\chi^2 = 577$ for 37
degrees of freedom.

It was pointed out by Irwin et al.\ (1999) that the trend in
the temperature profile found by the method of Markevitch et al.\ (1996)
differed from that found by other PSF-correction techniques from various
authors (e.g., Ikebe 1995; Fujita et al.\ 1996; Kikuchi et al.\ 1999).
Of the 28 clusters analyzed using the method of Markevitch et al.\ (1996)
that did not show very strong evidence for a merger, 22 (14) of them
were inconsistent with a constant temperature profile at the 70\% (90\%)
probability level. Conversely, clusters analyzed
using the method of Ikebe (1995), Fujita et al.\ (1996), or
Kikuchi et al.\ (1999) were found to have basically flat temperature profiles,
with all 11 clusters consistent with a constant temperature.

Of particular interest are clusters analyzed by more than one method, or
clusters whose temperature is low enough that the PSF of {\it ASCA} does
not influence the temperature profile significantly (for clusters below 5 keV).
A399, A401, MKW3S, A1795, and A496 are examples where the method of
Markevitch et al.\ (1996) leads to different trends in the radial
temperature profiles than with other methods (Fujita et al.\ 1996;
Kikuchi et al.\ 1996; Ohashi et al.\ 1997; Dupke \& White 1999). The
latter two are also in our sample, and show no evidence for a decline
in temperature outside of the cooling region (Figure~\ref{fig:temp_profiles}).

Recently, White (1999) has determined the temperature profiles of a large
number of clusters observed with {\it ASCA} using the PSF-correction
technique outlined in White \& Buote (1999). He found that 90\% of the
98 clusters in his sample were consistent with isothermality at the 3$\sigma$
confidence level. On a cluster by cluster comparison of the 11 clusters
common in both samples, we find our temperature profiles are in excellent
agreement with those of White (1999), with only two exceptions.  Our
temperature value of the outermost bin of A3266 is $\sim$2 keV higher than
the value of White (1999). However, the 1$\sigma$ error bars just touch
so this discrepancy is not at a high significance. Also, our temperature
profile of A496 differs somewhat from his. This difference occurs
around 4$^{\prime}$ and is significant at $\sim$$2\sigma$. However, given
that we have 44 temperature values for the 11 clusters, it is not
unreasonable that two of our values differ by $2\sigma$. In fact, this
is the number of $2\sigma$ discrepancies one would expect from a statistical
point of view.

Given the conflicting results regarding this topic, it is worthwhile to
examine the PSF-correction techniques of the two largest studies, namely
those using the techniques of Markevitch et al.\ (1996; hereafter method M)
and White \& Buote (1999; hereafter method WB). Both techniques assume a
spatial distribution for the cluster emission; method M assumes the
emissivity profile obtained from {\it ROSAT} PSPC data for the cluster
in question, while method WB assumes the spatial distribution derived
from a maximum-likelihood deconvolution of the {\it ASCA} image. Method
M creates simulated events drawn from this spatial distribution and
from an initial guess for the cluster spectrum, and convolves the
events with the spatially-variable PSF. The input cluster spectrum is
varied until the simulated data matches the actual data.
Method WB ray-traces events drawn from the spatial distribution through
the telescope optics and attempts to find the most likely association
between events in the deconvolved and convolved planes on a PI energy bin
by PI energy bin basis. Once an energy has been assigned to each event
in the deconvolved plane, the event list can be analyzed via standard
spectral-fitting procedures.

Both methods have their advantages and
disadvantages. Whereas method WB employs only a spatially-invariant PSF,
method M uses a PSF that varies with position. On the other hand, method M
is reliant on {\it ROSAT} data to determine the emissivity profile, while
method WB is not. It is not clear if the emissivity profile derived from
the 0.2--2.0 keV {\it ROSAT} band is appropriate for use over the
{\it ASCA} bandpass. The only way to determine which of these disadvantages
are leading to incorrect results is to create simulated data
with a known temperature and surface brightness profile and convolve it
with the response of the instrument, and then see if the PSF-correction
technique can recover the input temperature and surface brightness profiles.
White \& Buote (1999) have tested their code on simulated cluster data with
a giant cooling flow model, a medium cooling flow model,
an isothermal profile, and a profile that decreases by a factor of two from
the center out to 20$^{\prime}$, as well as data of varying signal-to-noise
ratios. In all cases, the original temperature and surface brightness
profiles are recovered. Conversely, method M has not been as extensively
tested on a variety of simulated temperature distributions and signal-to-noise
ratios.

As a final note, it should be noted that Markevitch et al.\ (1998) used
a cooling flow model to fit the innermost bin of cooling flow clusters,
whereas White (1999) used a single component thermal model. This will
naturally lead to a lower value for White (1999) for the innermost bin
(typically $\sim5\%$ of the virial radius).
However, White (1999) claims that correcting for the cooling gas leads to an
ambient core temperature within the innermost bin consistent with the outer
regions of the cluster. Again, this is inconsistent with the results of
Markevitch et al.\ (1998) who find a value for the ambient core temperature
greater than the rest of the cluster.

\begin{figure*}[htb]
\vskip4.6truein
\includegraphics{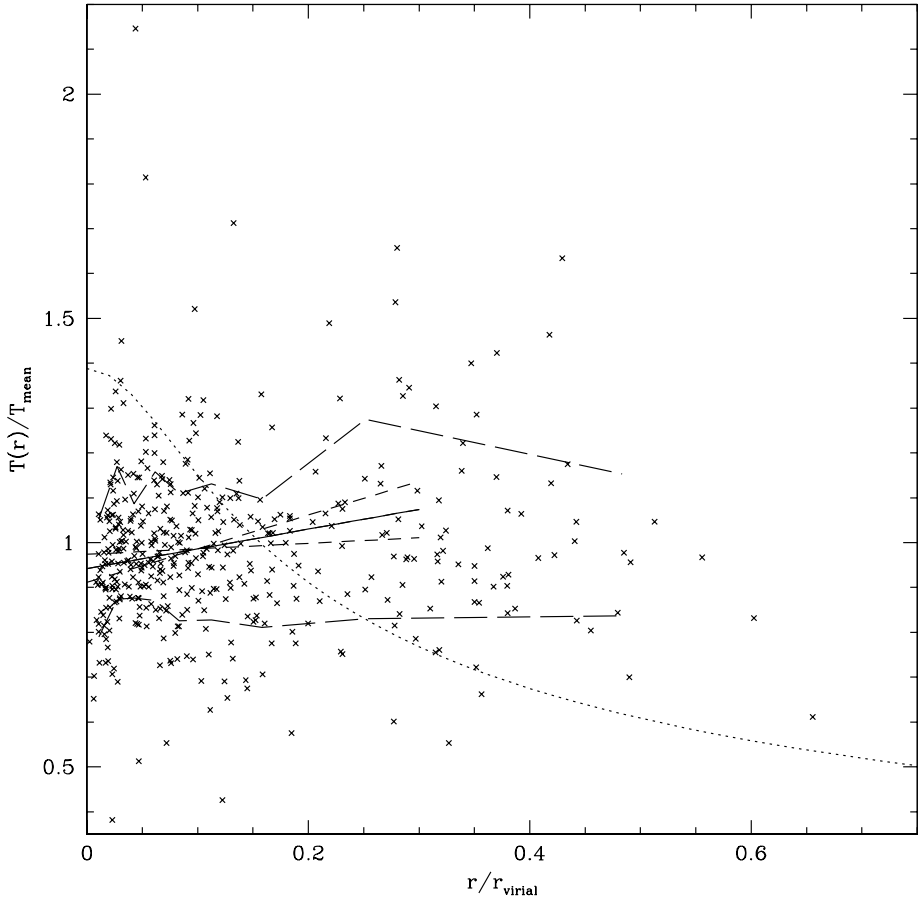}
\caption[summary]{
Summary of cluster temperature profiles to date. The normalized temperature
profiles of 98 clusters observed by White (1999) as a function of the
virial radius are shown as crosses. The long dotted lines enclose the middle
68\% of the data points. Also shown is the Markevitch et al.\ (1998)
result (dotted line). The solid line with the accompanying dashed lines
represents the composite temperature profile presented in this paper along
with the 90\%  errors in the slope of the best-fit line.
\label{fig:summary}}
\end{figure*}

\subsection{Temperature Profiles Determined With {\it ROSAT}} \label{ssec:rosat}

Although the limited bandpass of the {\it ROSAT} PSPC (0.2-2.4 keV) precludes
tight constraints to be put on the temperatures of hot clusters, large (factor
of two) temperature changes should be detectable with large enough
signal-to-noise ratios. Unfortunately, most clusters observed with {\it ROSAT}
did not have good enough statistics to accomplish this on an individual basis.
To circumvent this problem, Irwin et al.\ (1999) averaged together the
radial color profiles (ratio of counts in various bands covering the
{\it ROSAT} PSPC bandpass) of 26 clusters observed with the PSPC. If large-scale
deviations from isothermality were common in clusters, such a feature
lost in the noise for an individual cluster would become apparent when
the clusters were added together. Although a drop in temperature 
was found in the center of cooling flow clusters (indicating that the
method could indeed detect changes in temperatures even for hot clusters),
the temperature profiles were flat outside of the cooling region out to
35\% of the virial radius. It was found that a 20\% temperature drop within
35\% of the virial radius was ruled out at the 99\% confidence level.
This is in agreement with the {\it BeppoSAX} data presented here, where
a decline in temperature of 14\% out to 30\% of the virial radius is ruled
out at the 99\% confidence level.

\subsection{Temperature Profiles Determined With {\it BeppoSAX}}
\label{ssec:bepposax}

Several of the clusters in our sample have been analyzed by other authors:
A2199 (Kaastra et al.\ 1998), A2029 (Molendi \& De Grandi 1999),
A2319 (Molendi et al.\ 1999), and A3266 (De Grandi \& Molendi 1999).
The profile of A2199 is fully consistent with ours. For the other three
clusters, somewhat different steps were taken in the data reduction process
than what we did. Whereas we excluded data below 3.0 keV, other authors
have included data down to 2.0 keV. In addition, they have frozen the
redshift whereas we have let it be a free parameter. We have re-analyzed
the {\it BeppoSAX} data for these three clusters, this time including data
down to 2.0 keV and freezing the redshift. Although inclusion of channels
below 3.0 keV lowered the derived temperatures somewhat (see
\S~\ref{ssec:temp_global}), it did not change the overall shape of the
profiles.

Our new profile of A2319 was consistent with that of Molendi et al.\ (1999).
The profiles of A2029 and A3266 differed somewhat from
Molendi \& De Grandi (1999) and De Grandi \& Molendi (1999). However,
the differences occurred only in the 2$^{\prime}-4^{\prime}$ bins, with
our 1$\sigma$ error bars overlapping their 1$\sigma$ error bars for the
other three spatial bins. For both A2029 and A3266 our 2$^{\prime}-4^{\prime}$
temperature was $\sim$1.5 keV lower than those obtained by
Molendi \& De Grandi (1999) and De Grandi \& Molendi (1999), and the difference
was significant at the 3.3$\sigma$ and 1.9$\sigma$ confidence levels
for A2029 and A3266, respectively. The cause
of this discrepancy may be the choice of the assumed surface brightness profile,
especially for the case of A2029, which possesses a large cooling flow.
As mentioned in \S~\ref{sec:sample}, we have used the double-beta model
profile obtained by Mohr et al.\ (1999) from {\it ROSAT} PSPC data. It is
not stated where the other authors obtained their assumed surface brightness
profiles. However, the agreement between the other three spatial
bins (especially the first and fourth spatial bins) is encouraging,
indicating a flat temperature profile out to $9^{\prime}$ for A2029
and A2319, and a modestly decreasing profile for A3266. It should be
noted that Molendi \& De Grandi (1999) and De Grandi \& Molendi (1999)
find significant temperature drops in A2029 and A3266, respectively,
at very large radii, where we have truncated our profiles because of
the presence of the strongback, and the increasing asymmetry of the
PSF at larger radii.

\section{A Summary of Cluster Temperature Profiles} \label{sec:summmary}

Given the effort put forth in determining temperature profiles for a large
sample of clusters
in recent years it is worthwhile to summarize the major contributions to
this subject: Markevitch et al.\ (1998; and references within) and White (1999)
with {\it ASCA} data, Irwin et al.\ (1999) with  {\it ROSAT} PSPC data, and
this current study with {\it BeppoSAX} data. The combined results are shown
in Figure~\ref{fig:summary}. The 98 temperature profiles of White (1999)
(shown as crosses) have been normalized to the global temperature for each
individual cluster and scaled in units of the virial radius. The long dashed
lines enclose the middle 68\% of the data points. The error bars are
typically rather large for each data point, and have been excluded for
clarity. The temperature profile derived
by Markevitch et al.\ (1998) is shown as a dotted line, and represents
a fit with polytropic index of $\gamma=1.24$. The results presented in this
paper is shown as a solid line. The {\it ROSAT} PSPC result is not
shown but is very similar to best-fit line derived here, only somewhat
less constrained.

The results of White (1999), Irwin et al.\ (1999), and this study all point
to the same general conclusion: outside of the cooling region but inside
30\% of the virial radius there appears to be no decline in the temperature
profile. White (1999) extends this result out to 45\% of the virial radius,
although it should be noted that large drops in the temperature have been
found in A2029 and A3266 with {\it BeppoSAX} data (Molendi \& De Grandi 1999;
De Grandi \& Molendi 1999) in this region. Outside of 45\% of the virial
radius White (1999) finds evidence for a decline in temperature although
the statistics are sparse in this region. This drop is not surprising though
since nearly all cluster simulations show a decline in temperature at large
radii.

\acknowledgments
JAI thanks R. Dupke for many useful comments and conversations. We thank
D. White for kindly providing us with his {\it ASCA} temperature profiles, and
also the anonymous referee for many insightful comments and
suggestions to improve the paper.
This research has made use of data obtained through the High Energy
Astrophysics Science Archive Research Center Online Service,
provided by the NASA/Goddard Space Flight Center, and also the
{\it BeppoSAX} Science Data Center.
This work has been supported by {\it Chandra} Fellowship grant PF9-10009,
awarded through the {\it Chandra} Science Center. The {\it Chandra} Science
Center is operated by the Smithsonian Astrophysical Observatory for NASA
under contract NAS8-39073.

\end{document}